\documentclass[aps, prd, twocolumn, amsmath, floats,floatfix, superscriptaddress, nofootinbib,showpacs]{revtex4}

\usepackage{amsmath}
\usepackage{bm}
\usepackage{array}
\usepackage{ifpdf}

\ifpdf 
  \usepackage[pdftex]{graphicx}
  \graphicspath{{figures/pdf/}}
  \DeclareGraphicsExtensions{.pdf}
\else
  \usepackage[dvips]{graphicx}
  \usepackage{graphicx}
  \usepackage{pstricks}
  \usepackage{pst-node}
  \usepackage{pst-blur}
  \graphicspath{{figures/eps/}}
  \DeclareGraphicsExtensions{.eps}
  \definecolor{Pink}{rgb}{1.,0.75,0.8}
  \definecolor{cyan2}{cmyk}{0.40,0,0,0}
\fi
\def\diag{\operatorname{diag}}

\def\p0{\partial_0}

\usepackage{tikz}
\usepackage{amsmath}
\usepackage{amssymb}
\usepackage{listings}
\usepackage{mathtools}
\usepackage{subfigure}

\begin{document}

\title{Measuring the black hole spin direction in 3D Cartesian numerical relativity simulations}

\author{Vassilios Mewes}
\affiliation{Departamento de
  Astronom\'{\i}a y Astrof\'{\i}sica, Universitat de Val\`encia,
  Dr. Moliner 50, 46100, Burjassot (Val\`encia), Spain}

\author{Jos\'e A. Font}
\affiliation{Departamento de
  Astronom\'{\i}a y Astrof\'{\i}sica, Universitat de Val\`encia,
  Dr. Moliner 50, 46100, Burjassot (Val\`encia), Spain}
\affiliation{Observatori Astron\`omic, Universitat de Val\`encia, C/ Catedr\'atico 
  Jos\'e Beltr\'an 2, 46980, Paterna (Val\`encia), Spain} 

\author{Pedro J. Montero} 
\affiliation{Max-Planck-Institute f{\"u}r Astrophysik, Karl-Schwarzschild-Str. 1, 85748, Garching bei M{\"u}nchen, Germany}
  
\begin{abstract}
We show that the so-called flat-space rotational Killing vector method
for measuring the Cartesian components of a black hole spin can be
derived from the surface integral of Weinberg's pseudotensor over the
apparent horizon surface when using Gaussian normal coordinates in
the integration. Moreover, the integration of the pseudotensor in this
gauge yields the Komar angular momentum integral in a foliation adapted
to the axisymmetry of the spacetime. As a result, the method does
not explicitly depend on the evolved lapse $\alpha$ and shift $\beta^i$
on the respective timeslice, as they are fixed to Gaussian normal
coordinates, while leaving the coordinate labels of the spatial
metric $\gamma_{ij}$ and the extrinsic curvature $K_{ij}$ unchanged.
Such gauge fixing endows the method with coordinate invariance,
which is not present in integral expressions using Weinberg's
pseudotensor, as they normally rely on the explicit use of
Cartesian coordinates. 
\end{abstract}

\pacs{
04.25.Dm, 
95.30.Sf, 
97.60.Lf  
}

\maketitle

\section{Introduction}
\label{sec:introduction}

Since the binary black hole (BBH) merger breakthrough simulations of 
about a decade ago~\cite{Pretorius2005,Baker2006,Campanelli2006}, 
ever-growing computational resources and advances in the numerical 
methods used to simulate these systems have made the exploration 
of the vast initial parameter space possible (see e.g.~\cite{Hinder2010} 
and references therein for a recent overview of the status of BBH
simulations). The initial parameters of BBH simulations are the BH
mass ratio and the six components of their initial spin vectors. The
investigation of these initial parameters has led to significant
discoveries, as the occurrence of the orbital
hang-up~\cite{Campanelli2006a} and the presence of the so-called 
super-kicks where the final BH is displaced from  the orbital plane 
after its formation~\cite{Gonzalez2009}.

The BH spin and in particular its orientation may also play a
non-negligible role in non-vacuum  spacetimes involving BHs surrounded 
by matter as, e.g.~in the form of accretion disks, a situation
commonly encountered in binary neutron star merger
simulations. Recently,  we have performed numerical relativity
simulations of {\it tilted} self-gravitating accretion disks around
BHs, investigating the precession and nutation the BH undergoes as it 
accretes mass and angular momentum from the torus~\cite{Mewes2015}. 
In order to carry out a quantitative analysis of such kind of
simulations, it is obviously necessary to measure both the
magnitude of the BH spin and also its direction in space. 
Such a study inspired the work we present in this paper.

One of the standard methods in numerical relativity to measure the
magnitude of the angular momentum of the BH horizon is described
in~\cite{Dreyer2003}. This method is based on the so-called isolated 
horizon formalism~\cite{Ashtekar1999} and the generalization to
dynamical horizons~\cite{Ashtekar2003}. In this approach the BH spin
is calculated by performing the following surface integral on the apparent horizon (AH) of the BH
\begin{eqnarray}
\label{eq:qlm_spin_integral}
 J_{\mathrm{AH}} = \frac{1}{8 \pi} \int_S \left( \psi^a R^b K_{a b} \right) dS,
\end{eqnarray}
where $\psi^a$ is an approximate rotational Killing vector on the
horizon that has to be determined numerically (see ~\cite{Dreyer2003}
for a method of finding $\psi^a$), $R^b$ is the outward pointing unit
vector normal to the horizon, $K_{a b}$ is the extrinsic curvature on
the horizon surface and $dS$ is the surface element. 
This method does not, however, give the direction of the BH spin in
the 3D Cartesian reference frame of the computational grid of a
numerical relativity simulation. 

The direction of the BH spin in numerical relativity is commonly
measured by the approach suggested by~\cite{Campanelli2007}. In this 
approach the BH spin direction is simply defined as the Euclidean unit 
vector tangent to the coordinate line joining the two poles on the
horizon (i.e.~the two points where the axially symmetric vector $\psi^a$
vanishes). The approximate Killing vector field $\psi^a$ on the
horizon is obtained numerically using spherical-polar coordinates and
the accuracy in the spin direction is typically about a few angular
grid zones. This definition of the spin vector reproduces the
Bowen-York spin parameter on the initial slice and gives satisfactory
results as long as the BH horizon does not become too
distorted. Moreover,~\cite{Campanelli2007} present another method
for finding the spin magnitude and direction, using flat-space coordinate
rotational Killing vectors to calculate the Cartesian components of the BH
spin and its magnitude from the Euclidean norm of the resulting vector. 
The flat-space Killing vector method has the practical advantage that the vector $\psi^a$  
used in the surface integral (\ref{eq:qlm_spin_integral}) is given 
analytically and is constant, therefore it does not have to be found numerically on each timeslice. 

In this paper we show how the flat-space Killing vector method can be 
derived by performing a surface integral of Weinberg's energy-momentum
pseudotensor~\cite{Weinberg1972}. 
By using the 3+1 split of spacetime and Gaussian coordinates, it is
possible to express the angular momentum 
of a given volume using Weinberg's energy-momentum pseudotensor in a simple form that allows for a
straightforward calculation of the spin vector of the BH horizon. 
Weinberg's energy-momentum pseudotensor is a symmetric pseudotensor derived by writing 
Einstein's equations using a  coordinate system that is
quasi-Minkowskian, i.e.~with the four-dimensional metric $g_{\mu\nu}$
approaching the Minkowski metric $\eta_{\mu \nu}$ at infinity. Although it is not generally
covariant, the pseudotensor is Lorentz covariant, and with the appropriate choice of
coordinates it provides a measure of the total angular momentum of
the system. In the
following Greek indices run from 0 to 3 while
Latin indices run from 1 to 3. We use geometrized units ($G=c=1$) 
throughout.

\section{3+1 surface integral of Weinbergs's pseudotensor in Gaussian coordinates}
\label{sec:ang_mom_volume}

In this section, we first briefly review the calculation of the angular momentum contained in a volume using 
Weinberg's pseudotensor. Next, we express the resulting surface integral in terms of the 3+1 spacetime variables on 
a given timeslice. Finally, we show that by choosing Gaussian coordinates the 
integral reduces in complexity and is analytically equivalent to the flat-space rotational Killing vector method.

\subsection{Angular momentum with Weinberg's pseudotensor}
\label{sec:weinberg_construction}

Weinberg's energy-momentum pseudotensor is obtained by writing the Einstein equations in a 
coordinate system that is quasi-Minkowskian in Cartesian coordinates, so that the metric $g_{\mu \nu}$ 
approaches the Cartesian Minkowski metric $\eta_{\mu \nu}=\diag(-1,1,1,1)$ at infinity as follows
\begin{eqnarray}
g_{\mu \nu} = \eta_{\mu \nu} + h_{\mu \nu}\,,
\label{eq:metric_split}
\end{eqnarray}
where $h_{\mu \nu}$ does not necessarily have to be small
everywhere. Then, by writing the Einstein equations in parts linear in $h_{\mu \nu}$, one 
arrives at an energy-momentum pseudotensor $ \tau^{\mu \nu}$, which is the total energy-momentum ``tensor'' of
the matter fields, $T_{\lambda \kappa}$, and of the gravitational field, $t_{\lambda \kappa}$,
\begin{eqnarray}
\label{eq:weinberg_pseudotensor}
 \tau^{\mu \nu} = \eta^{\mu \lambda} \eta^{\nu \kappa}\left[T_{\lambda \kappa} + t_{\lambda \kappa} \right] = \frac{1}{8\pi} \frac{\partial}{\partial x^{\sigma}} Q^{\sigma \mu
   \nu}_{\,\,\,\,\,\,\,\,\,\,},
\end{eqnarray}
where $Q^{\sigma \mu \nu}$ is the superpotential given by
\begin{eqnarray}
\label{eq:weinberg_superpotential}
 Q^{\sigma \mu \nu} &=& \frac{1}{2} \left( \frac{\partial h^{\lambda}_{\lambda}}{\partial x_{\mu}} \eta^{\sigma \nu}- 
 \frac{\partial h^{\lambda}_{\lambda}}{\partial x_{\sigma}} \eta^{\mu \nu}-
 \frac{\partial h^{\lambda \mu}}{\partial x^{\lambda}} \eta^{\sigma \nu} \right. \nonumber \\
 &+& \left. \frac{\partial h^{\lambda \sigma}}{\partial x^{\lambda}} \eta^{\mu \nu}+
 \frac{\partial h^{\mu \nu}}{\partial x_{\sigma}}-
 \frac{\partial h^{\sigma \nu}}{\partial x_{\mu}} \right),
\end{eqnarray}
and indices of linearized quantities are raised and lowered with
$\eta_{\mu \nu}$.

Using the pseudotensor, the volume integrals giving the total four-momentum
of the volume are given by
\begin{eqnarray}
\label{eq:weinberg4_mom_volume}
 P^{\mu} = \int_V \tau^{0\mu} d^3 x = -\frac{1}{8\pi} \int_V \left(\frac{\partial Q^{i 0
       \mu}}{\partial x^i}\right) d^3 x.
\end{eqnarray}

Furthermore, the pseudotensor $\tau^{\mu \nu}$ defined by
Eq.~(\ref{eq:weinberg_pseudotensor}) 
is symmetric, which allows one to use it to calculate the total angular momentum in a 
volume $V$ using the following volume integral:
\begin{eqnarray}
\label{eq:weinberg_ang_mom_volume_integral}
 J^{\mu \nu} &=& \int_V \left(x^{\mu} \tau^{0 \nu} - x^{\nu} \tau^{0 \mu} \right) d^3 x \nonumber \\
 &=& -\frac{1}{8 \pi} \int_V \left(x^{\mu} \frac{\partial Q^{i 0 \nu}}{\partial x^i}- x^{\nu} \frac{\partial Q^{i 0 \mu}}{\partial x^i}\right) d^3 x.
\end{eqnarray}

As Weinberg remarks, the physically interesting Cartesian components of the angular momentum contained in the volume are
\begin{eqnarray}
\label{eq:physical_components_ang_mom}
 J_x \equiv J^{23},\,\,\,\, J_y \equiv J^{31},\,\,\,\, J_z \equiv J^{12}.
\end{eqnarray}

Using Gauss' law the volume integral can be transformed to the following surface integral over the bounding surface: 
\begin{eqnarray}
\label{eq:weinberg_ang_mom}
  J^{i j} &=& -\frac{1}{16 \pi} \iint_S \left( -x_i \frac{\partial h_{0 j}}{\partial x^k} +x_j \frac{\partial h_{0 i}}{\partial x^k} 
  + x_i \frac{\partial h_{j k}}{\partial t}
  \right.
  \nonumber \\
    &-& \left.  x_j \frac{\partial h_{i k}}{\partial t}  + h_{0 j} \delta_{k i} - h_{0 i} \delta_{k j} \right) n^k dS, 
\end{eqnarray}
where $n^i$ is the unit normal to the surface of integration and $dS$ the surface element. 

The convergence of the four-momentum volume integrals
(\ref{eq:weinberg4_mom_volume}) involving the pseudotensor $\tau^{\mu \nu}$ critically depends on the rate
at which the metric $g_{\mu \nu}$ approaches the Minkowski reference
metric at large distances. Given the following behaviour of $h_{\mu \nu}$ as $r
\rightarrow \infty$,
\begin{eqnarray}
\label{eq:weinberg_quasi_minkowskian}
\begin{gathered}
 h_{\mu \nu} = \mathcal{O} (r^{-1}), \\
\frac{\partial h_{\mu
     \nu}}{\partial x^{\sigma}} = \mathcal{O} (r^{-2}),\\
\frac{\partial^2 h_{\mu
     \nu}}{\partial x^{\sigma}\partial x^{\rho}}=\mathcal{O} (r^{-3}),
\end{gathered}
\end{eqnarray}
where $r=(x^2+y^2+z^2)^{\frac{1}{2}}$, it can be shown that the energy-momentum
``tensor'' of the gravitational field, $t_{\mu \nu}$, behaves at large distances as
\begin{equation}
 t_{\mu \nu} =  \mathcal{O} (r^{-4}),
\end{equation}
which in turn shows that the four-momentum volume integral 
(\ref{eq:weinberg4_mom_volume}) converges. The convergence of the
total angular momentum volume integral
(\ref{eq:weinberg_ang_mom_volume_integral}) and of the corresponding
surface integral (\ref{eq:weinberg_ang_mom}) is more problematic, due to the
appearance of $x^{\mu}$ in the volume integral. This is also
observed in the convergence properties of the integrals of the ADM
quantities \cite{Arnowitt2008}, where the surface integrals for the ADM
mass and linear momentum converge when imposing fall-off conditions
like those of Eq.~(\ref{eq:weinberg_quasi_minkowskian}), while the calculation of the ADM 
angular momentum generally requires stronger asymptotic fall-off
conditions \cite{Jaramillo2011}. We shall return to the issue of the
convergence of Eq.~(\ref{eq:weinberg_ang_mom}) after we have expressed it
in terms of the 3+1 variables and in Gaussian normal coordinates in
the next section.

\subsection{The angular momentum pseudotensor integral in Gaussian coordinates}
\label{sec:weinberg_gaussian}

We can express the total angular momentum given by
Eq.~(\ref{eq:weinberg_ang_mom}) in 
Gaussian normal coordinates (also called synchronous coordinates), 
which represent free-falling observers. We start by doing a 3+1
decomposition of the four-dimensional metric $g_{\mu\nu}$,
\begin{eqnarray}
 g_{\mu \nu} = \left( \begin{array}{c|c}
  -\alpha^2 + \beta_i \beta^i & \gamma_{i j} \beta^j \\
  \hline
  \gamma_{i j} \beta^j & \gamma_{i j}
 \end{array}\right),
\end{eqnarray}
where $\alpha$ is the lapse function, $\beta^i$ the shift vector, and
$\gamma_{i j}$ the spatial metric induced on the hypersurface. 
From the requirement that the
metric $g_{\mu\nu}$ approaches Cartesian Minkowski space at infinity
(\ref{eq:metric_split}), we see that 
\begin{eqnarray}
 h_{\mu \nu} = \left( \begin{array}{c|c}
  -\alpha^2 + \beta_i \beta^i +1 & \gamma_{i j} \beta^j \\
  \hline
  \gamma_{i j} \beta^j & \gamma_{i j} -\delta_{i j}
 \end{array}\right).
\end{eqnarray}

If we now express the angular momentum surface integral,
Eq.~(\ref{eq:weinberg_ang_mom}), in terms of the 3+1 variables we find
that $J^{i j}$ can be written as 
\begin{multline}
\label{eq:weinberg_3_1}
  J^{i j} = -\frac{1}{16 \pi} \iint_S \bigg( -x_i \frac{\partial (\gamma_{j m} \beta^m)}{\partial x^k} +x_j \frac{\partial (\gamma_{i m} \beta^m)}{\partial x^k} \\
  + x_i \frac{\partial (\gamma_{j k} -\delta_{j k})}{\partial t} - x_j \frac{\partial (\gamma_{i k} -\delta_{i k})}{\partial t}\\
  + \gamma_{j m} \beta^m \delta_{k i} - \gamma_{i m} \beta^m \delta_{k j} \bigg) n^k dS. 
\end{multline}

Moreover, in terms of the 3+1 variables, Gaussian coordinates are defined
by the following choice of the lapse and shift vector:
\begin{equation}
\alpha = 1, \,\,\,\,\,\,\, \beta^i = 0,
\end{equation}
so that $h_{00}=h_{0 i}=h_{i 0}=0$.
In this gauge, Eq.~(\ref{eq:weinberg_3_1}) considerably simplifies to
\begin{equation}\label{eq:weinberg_ang_mom_gaussian_coords}
 J^{i j} = -\frac{1}{16 \pi} \iint_S \left( x_i \frac{\partial \gamma_{j k}}{\partial t} - x_j \frac{\partial \gamma_{i k}}{\partial t} \right) n^k dS.
\end{equation}

We can now use the definition of the extrinsic curvature $K_{i j}$,
\begin{eqnarray}
 K_{i j} = - \frac{1}{2 \alpha} \left(\frac{\partial \gamma_{i j}}{\partial t} - \mathcal{L}_{\beta} \gamma_{i j} \right),
\end{eqnarray}
where $\mathcal{L}_{\beta}$ is the Lie derivative with respect to the
shift vector $\beta^i$, to see that the time derivative of the spatial metric
$\partial \gamma_{i j}/\partial t$ in Gaussian coordinates is simply
\begin{equation}\label{eq:3metric_dot}
 \frac{\partial \gamma_{i j}}{\partial t} = -2 K_{i j}.
\end{equation}

Substituting Eq.~(\ref{eq:3metric_dot}) in Eq.~(\ref{eq:weinberg_ang_mom_gaussian_coords}), we find that 
\begin{eqnarray}
 J^{i j} = \frac{1}{8 \pi} \iint_S \left( x_i K_{j k} - x_j K_{i k} \right) n^k dS. 
\end{eqnarray}

Finally, using Eq.~(\ref{eq:physical_components_ang_mom}), the three components of the Cartesian angular momentum
vector of a volume are given by
\begin{eqnarray}
\label{eq:components_ang_mom}
\begin{gathered}
 J_x = J^{23} = \frac{1}{8 \pi} \iint_S \left( y K_{3 k} - z K_{2 k} \right) n^k dS\,, \\
 J_y = J^{31} = \frac{1}{8 \pi} \iint_S \left( z K_{1 k} - x K_{3 k} \right) n^k dS\,, \\
 J_z = J^{12} = \frac{1}{8 \pi} \iint_S \left( x K_{2 k} - y K_{1 k} \right) n^k dS\,. 
\end{gathered}
\end{eqnarray}

Introducing the components of the three Cartesian Killing vectors of the rotational symmetry of
Minkowski space
\begin{eqnarray}
\label{eq:killing_vectors_minkowski}
\begin{gathered}
{\xi}_x = (0,-z ,y)\\
{\xi}_y = (z,0,-x)\\
{\xi}_z = (-y,x,0)
\end{gathered}
\end{eqnarray}
we can rewrite the surface integrals of the three Cartesian components
of the angular momentum in the following way:
\begin{eqnarray}
\label{eq:weinberg_killing_vector_integral}
J_i = \frac{1}{8\pi} \iint_S K_{jk} (\xi_i)^j n^k dS\, .
\end{eqnarray}
Thus, Weinberg's identification of the (2,3), (3,1) and (1,2) components
as being the physically interesting ones is now clearly seen from 
Eq.~(\ref{eq:weinberg_killing_vector_integral}), as it is the rotational
Killing vectors of Minkowski space that enter in the calculation of
the Cartesian components of the total angular momentum of the volume.

Note that this form of the angular momentum is remarkably similar to
that of the ADM angular momentum \cite{Jaramillo2011}:
\begin{equation}
\label{eq:ADM_ang_mom}
J_i = \frac{1}{8\pi} \lim_{r\to\infty} \iint_S \left( K_{jk} - K
  \gamma_{jk} \right) (\xi_i)^j n^k dS \, .
\end{equation}

If the integration is done over a sphere, the components of the surface normal $n^k$ are
given by
\begin{equation}
{n}^i = \left(\frac{x}{r}, \frac{y}{r}, \frac{z}{r}\right),
\end{equation}
so that $(\xi_i)^j$ and $n^k$ are orthogonal vectors,
\begin{equation}
 \gamma_{jk} (\xi_i)^j n^k = (\xi_i)^k n_k = 0 \,\,\,\,\,\forall\,i .
\end{equation}
Therefore, the part of the integral containing the trace of $K_{ij}$ in Eq.~(\ref{eq:ADM_ang_mom})
vanishes for spherical surfaces and therefore equations
(\ref{eq:weinberg_killing_vector_integral}) and (\ref{eq:ADM_ang_mom})
are identical. We have thus shown that by using Weinberg's
pseudotensor in Gaussian coordinates we obtain the total ADM angular
momentum evaluated at spatial infinity, when the integration surface
is a sphere. We might still need to impose a stricter asymptotic
behaviour than the asymptotic Euclidean flatness of
\cite{Jaramillo2011} (for instance the quasi-isotropic or asymptotic
maximal gauge), but as~\cite{Gourgoulhon2012} noted, the
$K_{jk} (\xi_i)^j n^k$ part of Eq.~(\ref{eq:ADM_ang_mom}) converges in practice.
We are, however, interested in evaluating
Eq.~(\ref{eq:weinberg_killing_vector_integral}) quasi-locally, that is, 
associated with finite 2-surfaces (in our actual applications, these will be apparent horizons of black 
holes~\cite{Mewes2015}).

For an axisymmetric spacetime, the angular momentum can be calculated
via the so-called Komar angular momentum \cite{Komar1959}, which is defined
as (following again the notation of \cite{Jaramillo2011,Gourgoulhon2012}):
\begin{equation}
\label{eq:komar_ang_mom}
J_K = \frac{1}{16\pi} \iint_S \nabla^{\mu} \phi^{\nu} dS_{\mu \nu},
\end{equation}
where $\phi^{\nu}$ is the axial Killing vector. Note the extra factor
of 2 in the denominator, known as Komar's anomalous factor
\cite{Katz1985}. 
The Komar angular momentum integral does not  have to be evaluated at
spatial infinity, but is valid for every surface. In \cite{Jaramillo2011,Gourgoulhon2012} it is shown that using a
slicing adapted to the axisymmetry of the spacetime, and expressing
Eq.~(\ref{eq:komar_ang_mom}) in terms of the 3+1 variables, the Komar
angular momentum becomes
\begin{equation}
\label{eq:komar_ang_mom_3+1}
 J_K = \frac{1}{8\pi} \iint_S K_{ij} \phi^i n^k dS.
\end{equation}
In \cite{Gourgoulhon2012} the above integral is evaluated for a Kerr
BH in spherical Boyer-Lindquist coordinates, and the angular momentum
is found to be $J_K=Ma$, as expected, where $M$ and $a$ are the black hole mass
and spin parameter, respectively. As the two integrals
(\ref{eq:weinberg_killing_vector_integral}) and
(\ref{eq:komar_ang_mom_3+1}) have exactly the same structure, and the latter
is coordinate (but not foliation) invariant, we arrive at the conclusion
that the introduction of Gaussian coordinates has led to a
coordinate invariant expression for the angular momentum derived from
Weinberg's pseudotensor, namely the Komar angular momentum. Note
the absence of the anomalous factor of 2 in our final expression 
(\ref{eq:weinberg_killing_vector_integral}). It therefore seems that
it is possible to relax the restriction of using Cartesian coordinates in calculations
involving Weinberg's pseudotensor. 

\subsection{Measuring the angular momentum in numerical relativity simulations}

It is easy to check that not only the choice of Gaussian coordinates
simplifies the calculation of the total angular momentum via
Weinberg's pseudotensor, but also that it makes straightforward the 
implementation of the above expressions in a numerical relativity
3D Cartesian code based on the 3+1 decomposition. For instance, if using the widely
adopted BSSN formulation~\cite{Nakamura1987,Shibata1995,Baumgarte1998}, 
the extrinsic curvature $K_{i j}$ of the spatial slices is
closely related to one of the evolved variables, namely the traceless
part of the conformally related extrinsic curvature. We note that in present-day numerical 
relativity simulations 
one does not typically use Gaussian coordinates for the
actual numerical evolutions. This has to do with the fact that
Gaussian coordinates can only be used in the close vicinity of a
spatial hypersurface, as the geodesics emanating from the
hypersurfaces will eventually cross and form caustics in a finite
time~\cite{Gourgoulhon2012}. Furthermore the foliation is not singularity-avoiding,
which means Gaussian coordinates are unsuitable for the numerical evolution of
spacetimes containing curvature singularities. 
Instead, the gauge conditions most commonly employed today
in numerical relativity belong to the family
of the so-called moving puncture gauges, which consist of the ``1+log'' 
condition for the lapse function~\cite{Bona1995} and the Gamma
driver condition for the shift vector~\cite{Alcubierre2003}. However,
one can use the numerical solution for the extrinsic curvature
$K_{ij}$ in Eq.~(\ref{eq:weinberg_killing_vector_integral}) due to the freedom to
 choose any gauge for calculations done on each timeslice.

In addition, Eq.~(\ref{eq:weinberg_killing_vector_integral}) is actually equivalent to the method
proposed by~\cite{Campanelli2007} for the calculation of  the angular momentum of a 
volume using flat-space coordinate rotational Killing vectors
(cf.~Eq.~(\ref{eq:killing_vectors_minkowski})).  To see this, consider the
definition of the Killing vectors in Cartesian coordinates
given by~\cite{Campanelli2007}:
\begin{eqnarray}
\label{eq:campanelli_killing_vectors}
\begin{gathered}
 \psi^a_x = [0,-(z-z_c),(y-y_c)],\\
 \psi^a_y = [(z-z_c),0,-(x-x_c)], \\
 \psi^a_z = [-(y-y_c),(x-x_c),0], 
\end{gathered}
\end{eqnarray}
where $(x_c, y_c, z_c)$ is the coordinate centroid of the apparent horizon, which has to be 
subtracted to avoid including contributions from a possible orbital angular momentum of the BH about the center of 
the computational grid in the calculation of its spin.
Upon substituting their flat-space coordinate rotational Killing
vectors into Eq.~(\ref{eq:qlm_spin_integral}), we find that
\begin{eqnarray}
\label{eq:coordspin}
\begin{gathered}
 J_x = \frac{1}{8 \pi} \iint_S \left( y K_{3 b} - z K_{2 b} \right) n^b dS\,, \\
 J_y = \frac{1}{8 \pi} \iint_S \left( z K_{1 b} - x K_{3 b} \right) n^b dS\,, \\
 J_z = \frac{1}{8 \pi} \iint_S \left( x K_{2 b} - y K_{1 b} \right)
 n^b dS\,,
\end{gathered}
\end{eqnarray}
where we have set $x_c=y_c=z_c=0$ for simplicity.
We see that the two sets of expressions for the Cartesian
components of the angular momentum vector of the AH, those from
Weinberg's pseudotensor evaluated 
in Gaussian coordinates and those from the flat space rotational
Killing vector method, are equivalent and equal to the Komar angular
momentum in an axisymmetric spacetime.

\section{Discussion}
\label{sec:gauge_depende}

As we have shown, the flat-space rotational Killing vector method of~\cite{Campanelli2007} can 
be derived from Weinberg's pseudotensor when using Gaussian coordinates. These coordinates have two
interesting properties that make them particularly useful for the evaluation of the angular momentum 
pseudotensor integral,  Eq.~(\ref{eq:weinberg_ang_mom}).
First, as we have shown, the complicated integral~(\ref{eq:weinberg_ang_mom}) reduces to the much
simpler expressions given by
Eq.~(\ref{eq:weinberg_killing_vector_integral}) and this final
expression is equal to the Komar angular momentum integral in a
foliation adapted to the axisymmetry of the system. 
As a result, one does not need the knowledge of the
shift vector and of its spatial derivatives on the surface of integration, 
which in practice would involve more quantities that one would need to
interpolate onto the horizon surface for the calculation of the
spin, thus also avoiding the numerical error associated with the
computation of the finite difference approximation to those spatial derivatives. 
Second, Gaussian coordinates trivially satisfy the 
necessary falloff conditions for the lapse and shift. 
Moreover, by using Gaussian coordinates we recover the ADM angular
momentum evaluated at spatial infinity, provided we use a
spherical surface of integration.

It is generally known that the various energy-momentum pseudotensors proposed in the literature are 
not covariant and care has to be taken when evaluating them in different coordinate systems and 
gauges. (See~\cite{Szabados2009} for a review on quasi-local mass and angular momentum in General Relativity, where the problems arising 
when using pseudotensors for the calculation of mass and angular momentum are also discussed.)
The derivation of Weinberg's pseudotensor relies crucially on the reference space 
being Cartesian Minkowski. In his textbook~\cite{Weinberg1972} Weinberg states that a spherical polar coordinate system would lead to a 
gravitational energy density concentrated at infinity. While being non-covariant is generally not desirable,
Weinberg's method is employed in a Cartesian grid, and Gaussian coordinates guarantee the correct asymptotic
behaviour of the lapse and shift, irrespective of the asymptotic behaviour the evolved lapse and shift may posses,
which are, as previously stated, not explicitly used in the
calculation of the AH spin on the respective timeslice.
Furthermore, we have shown that Gaussian coordinates transform the
pseudotensor angular momentum surface integral (\ref{eq:weinberg_3_1})
to the Komar angular momentum integral (\ref{eq:komar_ang_mom_3+1})
which is coordinate independent. The use of Gaussian coordinates
(as an explicit gauge-fixing) seems to therefore remove the coordinate
restrictions of the pseudotensor. 

When~\cite{Campanelli2007} introduced the flat-space 
rotational Killing vectors for the calculation
of the BH spin direction, the authors stated that they could not guarantee the
correct results for all times because the method
is not gauge invariant. However, as we have seen, 
such method can be derived from the integration of Weinberg's total
angular momentum pseudotensor over the apparent horizon surface when
using Gaussian normal coordinates in the integration. As a result, the
method does not depend on the evolving lapse and shift, as the gauge is fixed to the Gaussian normal coordinates on the respective timeslice.
We stress that the evolution of the lapse and shift during the free evolution of 
the spacetime does not enter the calculation, given the 
coordinates evolve in such a way that an AH is found 
at all times during the evolution, which is usually the case
in puncture evolutions with the BSSN system. There is a dependence on the gauge evolution via
the extrinsic curvature $K_{i j}$ that is interpolated onto the AH for the calculation of the spin
direction, but the same is true for the expression of the spin magnitude in Eq.~(\ref{eq:qlm_spin_integral}).

In \cite{Schnetter2006} the authors have shown that 
Eq.~(\ref{eq:qlm_spin_integral}) will give the spin magnitude provided
an approximate Killing vector can be found on the horizon and is gauge 
independent on the respective time-slice if the approximate Killing 
vector field $\psi^a$ is divergence-free. Here we have shown that both 
methods (i.e.~either via Weinberg's pseudotensor in Gaussian coordinates 
or via flat-space rotational Killing vectors) yield the Komar angular
momentum when the latter is expressed in a foliation adapted to the
axisymmetry. We note that the restriction to axisymmetry turns out in 
practice not to be a major weakness, as numerical relativity 
simulations repeatedly show that the remnants of binary black hole 
mergers and perturbed Kerr black holes typically settle down to
the axisymmetric Kerr solution quickly~\cite{Schnetter2006,Owen2009}.
Moreover, both methods provide a measure of the BH spin magnitude and 
direction that is not explicitly dependent on the lapse and the shift 
on the respective time-slice. As both methods use the fixed rotational 
Killing vectors of Minkowski space, they measure the spin contribution 
from the axisymmetry of the AH.

\begin{acknowledgments}
It is a pleasure to thank I.~Cordero-Carri\'on, D.~Hilditch, J.~L.~Jaramillo, R.~Lapiedra and
E.~Schnetter for useful discussions. V. Mewes would like to thank
Ewald M{\"u}ller and the MPA for the support during his visit. This work was supported  
by the Spanish Ministry of Economy and Competitiveness (MINECO) through grants AYA2010-21097-C03-01 and AYA2013-40979-P, by the Generalitat Valenciana (PROMETEOII-2014-069) and
by the Deutsche Forschungsgemeinschaft (DFG) through its Transregional Center SFB/TR7 ``Gravitational Wave Astronomy''.
 \end{acknowledgments}

\bibliographystyle{apsrev4-1-noeprint}
\bibliography{references}

\end{document}